\documentclass[preprint,amsmath,amssymb,floatfix]{revtex4-1}
\usepackage{graphicx}
\usepackage{dcolumn}
\usepackage{bm}
\usepackage[mathlines]{lineno}

\begin{document}
\title{Optimal vector beams maintaining robust intensity profile on propagation through turbulence}

\author{Priyanka Lochab}
\author{P. Senthilkumaran}
\author{Kedar Khare} 
\email{kedark@physics.iitd.ac.in}
\affiliation{Department of Physics, Indian Institute of Technology Delhi, New Delhi 110016 India}

\date{\today}

\begin{abstract}
We report experiments on propagation of scalar and vector optical beams through random phase screens mimicking turbulence and show that the intensity profile of the beam containing a C-point polarization singularity shows maximally robust behavior. This observation is explained in terms of the polarization and orbital angular momentum (OAM) diversity in the beam. The $l = 0$ and $l = 1$ OAM states whose vector combination leads to the C-point singularity are seen to produce complementary speckle intensity patterns with significant negative correlation on propagation through a random phase screen. This unique property of C-point singularity makes it superior to other inhomogeneous polarization states as demonstrated in our experiments. The results provide an important generic guideline for designing beams that can maintain optimally robust beam intensity profile on passing through random phase fluctuations and are expected to have a number of applications. 


\end{abstract}

\maketitle

\section{Introduction}
Robust propagation of light beams through randomly fluctuating media is required for important applications such as free space communication, LIDAR systems, laser guided defense systems, imaging through biological tissues, etc. When light beams encounter spatially varying refractive index, their intensity profile is degraded significantly beyond the nominal beam spreading due to diffraction effects. Refractive index variations as small as $10^{-5} - 10^{-6}$ in these situations are sufficient to make the beam profile speckled and hence not good enough for delivering sufficient energy within the diffraction limited spot. In such situations, a feedback system like adaptive optics is considered to be essential for faithful beam delivery. Our aim in this paper is to evolve an optimal beam engineering approach using the interplay between polarization and orbital angular momentum (OAM) states of light that is inherently best suited for sending light beams through turbulent media in a robust manner.  

Degradation of beam quality on propagation through turbulence has been investigated for various scalar beams like Bessel\cite{Birch:15,Eyyuboglu2007}, cos and cosh-Gaussian\cite{Eyyuboglu2011}, flat-topped \cite{1464-4258-8-6-008} and vortex carrying beams\cite{BOUCHAL2002155,Gu:13,Liu:11}. Effects of coherence\cite{Schulz:05} and polarization of the input laser beam on intensity scintillations have also been studied. Partially coherent and partially polarized beams are known to have lower scintillations compared to its fully coherent counterpart\cite{doi:10.1117/1.3090435,Borah:10,Ricklin:02,Zhang:17, KORKOVA20082342}. Phase and Stoke's phase singularities of optical beams have been shown to remain robust on propagation through a random medium\cite{Gbur:08, Cheng:09} over a certain distance, so that, they can be used as information carriers. In our opinion, intensity of the beam is simplest and easiest quantity to detect and hence laser beam designs which preserve intensity are much more valuable for practical applications.  

The use of inhomogeneously polarized vector beams for robust propagation through turbulence has been proposed before \cite{WANG20083617, Gu:09, Lochab:17, rakesh:2016}. An inhomogeneously polarized beam can be thought of as a superposition of two orthogonal and homogeneously polarized beams derived from the same source and each having a different amplitude and phase structure. Since the two orthogonally polarized states do not interfere, the overall intensity profile of an inhomogeneously polarized beam is the sum of the intensities of its two polarization components. If the two polarization states evolve into sufficiently diverse speckles on propagation through turbulence, the overall beam will have smoother intensity profile compared to its individual polarization components. Keeping this idea in mind, in the present work we explore various inhomogeneously polarized vector beams and provide an optimal beam designing principle.

We sent vector beams in various states of polarization through random phase screens to observe the far field speckle patterns and the 
intensity profile quality of the beams was evaluated quantitatively. We find that the OAM diversity in the orthogonal polarization components of the beam is important for obtaining speckle pattern diversity. In this context, polarization singularities are an important class of inhomogeneously polarized light as they naturally contain features of both polarization and OAM diversity \cite{DENNIS2002201,FREUND2002251}. The combination of $l = 0$ and $l = 1$ (or $l = -1$) OAM states in orthogonal polarizations leads to C-point singularities (e.g. lemon or star structures) whereas a combination of $l = 1$ and $l = -1$ OAM states leads to V-point singularities (e.g. radially and azimuthally polarized light). An important result of our study is that C-point singularity structures are the best at preserving intensity profile of the beam when compared with V-points singularities or any other non-singular inhomogeneously or homogeneously polarized beams.
 
\section{Speckle pattern diversity for orthogonal polarizations}
The transverse E-field profile of an optical beam nominally propagating in the $+z$ direction may be described as:
\begin{equation}
\vec{E}(x,y) = [\hat{e}_1 E_1 (x,y) + \hat{e}_2 E_2 (x,y)] \exp(i kz ),
\end{equation}
where $\exp(-i \omega t)$ time dependence has been assumed and the orthogonal unit vectors $\hat{e}_1$ and $\hat{e}_2$ may for example denote $\hat{x}$ and $\hat{y}$ polarization states or the right and left circularly polarized (RCP or LCP) states. When the functional form of the amplitude and phase of the two orthogonal polarization components $E_1 (x,y)$ and $E_2 (x,y)$ of the field are different, the resultant vector field has locally varying polarization state. As shown in Fig. 1 (a) when this field is incident on a random phase screen $p(x,y)$ and limited by an aperture $A(x,y)$ the resultant far-field diffraction intensity pattern may be described by:
\begin{align}
I_{Total} (u,v) &= I_1 (u,v) + I_2 (u,v) \nonumber \\&= | \mathcal{F} \{ E_1 (x,y) p(x,y) A(x,y) \} (u,v)|^2 + | \mathcal{F} \{ E_2 (x,y) p(x,y) A(x,y) \} (u,v)|^2.
\end{align}
Here $(u,v)$ denote the far-field transverse co-ordinates and $\mathcal{F}$ denotes the 2D Fourier transform operation. The above relation is approximately true as we are neglecting any cross-talk between the two orthogonal polarization components which is usually negligible unless we are dealing with thick scattering media. The single phase screen model for describing turbulent medium at a given instant of time is sufficient for the present purpose and is widely used for adaptive wavefront correction problems \cite{johansson}. The intensity patterns $I_1 (u,v)$ or $I_2 (u,v)$ for single polarization states become speckled as a result of the randomness in the phase screen $p(x,y)$ and the locations of intensity maxima and minima in the speckle patterns fluctuate randomly. As a result if a beam is to be received at a target or by a finite-sized detector, the amount of total energy delivered on a desired area changes unpredictably. In this context the diversity in the speckle pattern intensities $I_1 (u,v)$ and $I_2 (u,v)$ becomes an important factor in designing a robust beam. Clearly if $I_1$ and $I_2$ are highly correlated, there is no gain to be achieved in terms of reducing the beam fluctuations. The traditional literature on speckle has considered reduction in speckle contrast on addition of uncorrelated speckles \cite{dainty}. However if the two speckle patterns $I_1(u,v)$ and $I_2(u,v)$ are negatively correlated, the random fluctuations in the overall intensity can be reduced significantly as the locations of intensity maxima for one polarization component now coincide with the locations of intensity minima of the orthogonal polarization component. It is therefore important to understand the nature of the complex field profiles $ E_1 (x,y)$ and $ E_2 (x,y)$ that will produce speckles with different degrees of correlation and select the optimal beam profiles that produce negatively correlated or complementary speckle patterns.

\begin{figure}[ht]
\centering
\includegraphics[width=0.85\textwidth]{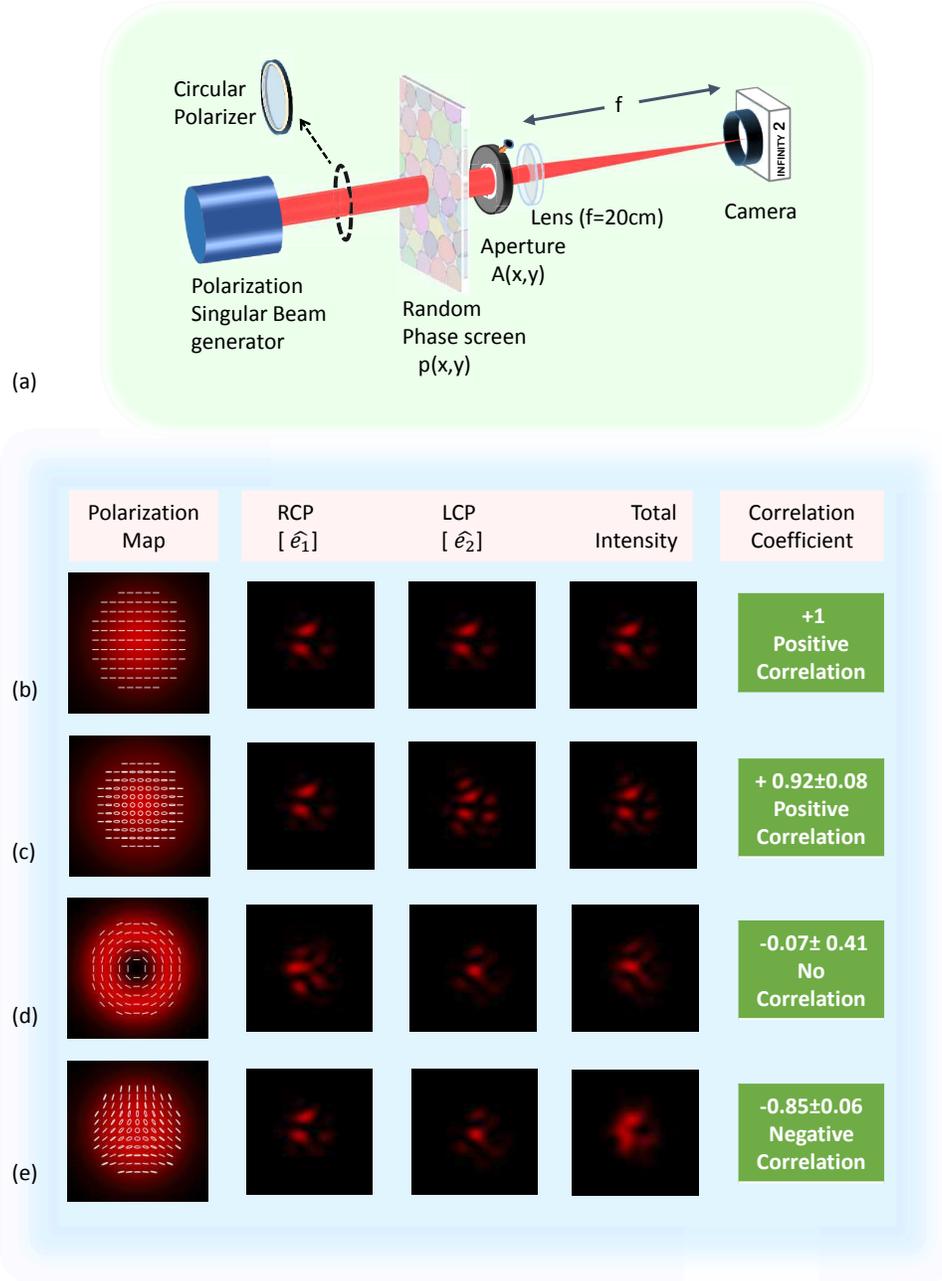}
\caption{(a) Experimental set-up for propagation of polarization singular beam through a random phase screen. Simulation results (b)-(e) show beams with different polarization structure and their corresponding far field intensity on passing through a random phase screen. The intensities for individual RCP and LCP components are also shown. The correlation coefficient between the RCP and LCP intensities is calculated over 50 random phase realizations}
\label{fig1}
\end{figure}

Figure \ref{fig1}(a) shows the basic experimental setup where an inhomogeneously polarized beam is incident on a random phase screen and the corresponding far-field diffraction intensity pattern is observed in the back focal plane of a lens. In Fig.s \ref{fig1}(b)-(e), four different combinations of $E_1(x,y)$ and $E_2(x,y)$ for the inhomogeneously polarized beam are shown. Here we have used $\hat{e}_1$ = RCP and $\hat{e}_2$ = LCP polarization states. The columns in Fig. 1 (b)-(e) depict the polarization state of the beam at the input, the intensities of $\hat{e}_1$ component, $\hat{e}_2$ component and total beam as observed on the camera respectively. The same realization of random phase screen is used here for all the illustrated cases. The last column shows the mean correlation coefficient values of the $\hat{e}_1$ and $\hat{e}_2$ component intensities. A set of $50$ independent realizations of the random phase screens was used for calculating the correlation coefficient. The correlation coefficient ($\rho_{II'}$) between two diffraction intensity patterns $I$ and $I'$ corresponding to the orthogonal polarizations $\hat{e}_1$ and $\hat{e}_2$ is calculated over diffraction limited spot size. It is defined as:
\begin{equation}
\rho_{II'}=\frac{E[(I-\mu_I)(I'-\mu_{I'})]}{\sigma_I \sigma_{I'}}
\end{equation}
where $E$ denotes the expectation, $\mu$ and $\sigma$ are the mean and standard deviation value.In Fig. \ref{fig1}(b) we show the case of a scalar or homogeneously polarized beam with equal energy in the $\hat{e}_1$ and $\hat{e}_2$ polarization components. The speckles produced by the two orthogonal polarization components in this case are identical and there is no advantage in terms of beam quality when the two intensities $I_1$ and $I_2$ are added. Next, we consider an inhomogeneously polarized beam obtained by using two different amplitude profiles for $E_1(x,y)$ and $E_2(x,y)$in the orthogonal polarizations as shown in Fig. \ref{fig1}(c). The amplitude profiles are selected as Gaussian for the $\hat{e}_1$-polarization and hollow Gaussian \cite{{Cai:06},{Song:99}} for the $\hat{e}_2$-polarization, keeping the phase identical so that the field incident on the random phase screen has the form:
\begin{equation}
\vec{E}(x,y)= A_o \exp(-r^2/w^2) [\hat{e_{1}} + (r^2/w^2) \hat{e_{2}}] \exp(i kz ),
\end{equation}
where $r = \sqrt{x^2 + y^2}$ is the radial coordinate in the $(x,y)$ plane, $w$ is the beam waist of the Gaussian mode and $A_o$ is the normalization constant. In this case $I_1$ and $I_2$ are found to be positively correlated with correlation coefficient equal to $0.92\pm 0.08 $. The sum intensity pattern therefore shows similar fluctuations as for the individual polarization components. Hence, any inhomogeneously polarized beam formed as a result of only amplitude diversity in the two polarization components would not be useful in maintaining robust intensity through turbulence. 

In the third case (Fig. \ref{fig1}(d)) we add phase diversity in the two polarization states in the form of different OAM states $l = 1$ and $l = -1$. Now, the two polarization states have phase singularities in the form $\exp(i \theta)$ and $\exp(-i \theta)$ respectively where $\theta = \arctan(y/x)$ is the polar angle in the transverse $(x,y)$ plane. The $l =+1$ and $l=-1$ states are embedded in the $\hat{e_1}$ and $\hat{e_2}$ polarization states respectively leading to an azimuthally polarized beam. It is observed that the correlation coefficient between $I_1$ and $I_2$ in this case is $-0.07 \pm 0.41 $ which indicates uncorrelated speckle patterns on an average. The speckle contrast in this case is seen to reduce for the total intensity as compared to the individual polarization components.  Finally in Fig. \ref{fig1}(e) we use the $l = 0$ and $l = 1$ OAM states with the $\hat{e_1}$ and $\hat{e_2}$ polarizations to generate a C-point singularity structure (a star in this case). We observe that the speckle patterns in the orthogonal polarization states have a negative correlation coefficient equal to $-0.85 \pm 0.06$. The sum intensity pattern in this case has significantly improved quality for the central spot compared to the $\hat{e_1}$ and $\hat{e_2}$  polarization states. This high negative correlation can be visually seen in the speckles corresponding to $\hat{e_1}$ and $\hat{e_2}$ polarization states as maxima in one polarization complements the minima of the other polarization, giving rise to a much more reduced speckle contrast in the  total beam. This complementarity property is unique to C-point structures and can be understood from the concept of the spiral phase quadrature transform \cite{Larkin:01, Khare:08}. Mathematically this complementarity is similar to that seen in sine-cosine quadrature components since spiral phase transform is the 2D analogue of the 1D Hilbert transform that connects sines and cosines. It has been shown that when beams carrying OAM states $l=0$ and $l=1$ illuminate an amplitude/phase aperture, the corresponding diffraction patterns are complementary in nature\cite{PhysRevLett.118.043903, Lochab:17, Sharma:15}. This complementary evolution of the two polarization states would keep the total intensity robust even in situations where turbulence is time-varying. The complementary speckle property of the C-point singularity beams is therefore an important feature that can be used as a general guideline for designing robust beams. The first two cases in Fig.s \ref{fig1} (b), (c) do not offer any advantage from the point of view of reducing speckle contrast of the beam. We therefore performed experiment on beams containing V-point and C-point polarization singularities.   

\section{Generation of polarization singular beams}
Polarization singularities are the local points in the optical field where some aspect of its polarization (e.g. handedness, orientation) is not defined. In 2D, the most general types of polarization singularities are classified as (i) C-points and (ii) V-points for elliptical and vector fields respectively. These two types of polarization singularities have intrinsically different OAM mixing and are known to evolve differently on propagation. In elliptically polarized optical fields, at C-points the state of polarization (SOP) is circular and hence the orientation angle ($\gamma$) of the polarization ellipse becomes undefined. There are three types of generic C-point singularities: lemon, monstar and star. In linearly polarized optical fields, the orientation of the electric field vector is undefined at V-points which are intensity null points. Radially and azimuthally polarized beams form a subset of V-point singularities. 

There are various experimental ways to generate polarization singular beams. These involve use of spatially-variable retardation plates\cite{MACHAVARIANI2008732}, birefringent crystals\cite{PhysRevLett.95.253901}, Wollaston prisms\cite{2040-8986-12-3-035406,Kurzynowski:12} and mixing of different OAM states in orthogonal polarizations using Spatial light modulator(SLM)\cite{1367-2630-9-3-078} or an interferometric arrangement\cite{Aadhi:16,BROWN201181}. 

\begin{figure}[h]
\centering
\includegraphics[width=0.75\textwidth]{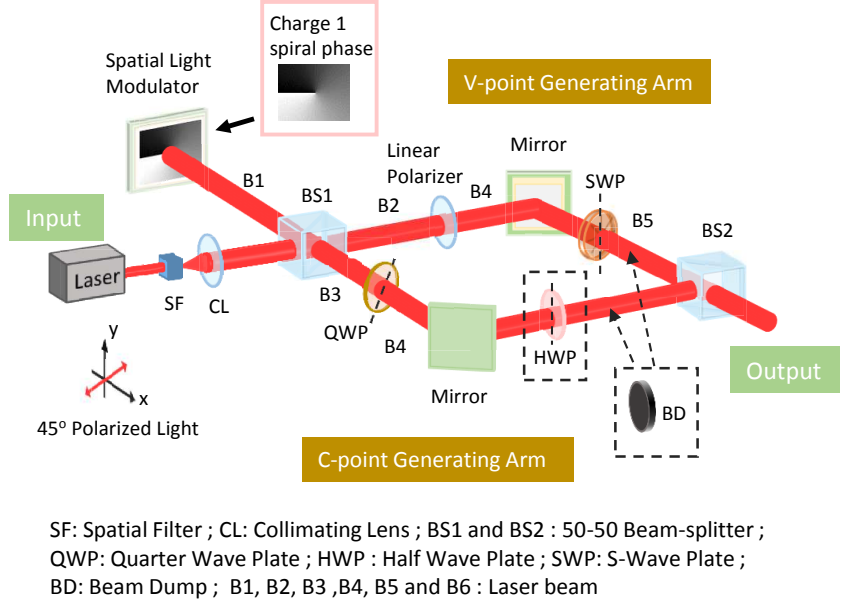}
\caption{Polarization singular beam generator}
\label{fig2}
\end{figure}

The setup used for the polarization singular beam generator is shown in Fig. \ref{fig2}. The input to the beam generator is a 45-degree polarized laser beam ($\lambda=650$ nm) which is incident on the beam-splitter (BS1). The output beam obtained at beam-splitter (BS2) constitutes the generated polarization singular beam. The two arms of the set-up are the V-point and C-point generating arms respectively. A beam dump(BD) is used in either one of the arms in order to choose the required output beam.  

Laser beam B1 is reflected from the Spatial Light Modulator (SLM)(Make: Holoeye, Model: LETO) displaying a charge-1 spiral phase pattern. SLMs are sensitive to only p-polarized light and act like a plane mirror for the s-polarized light. So the resultant light beam on reflection (Beam B3) from SLM is described as:
\begin{equation}
\vec{E}(x,y)= \psi(x,y)\hat{x} + e^{i\theta(x,y)}\psi(x,y)\hat{y}
\end{equation}
where $\psi$ is the beam profile of laser. It contains a polarization singularity dipole structure in the form of a lemon and star separated by a L-line. For generation of polarization singular beam carrying a single lemon or star, a coordinate transformation is done from rectangular to circular basis. Thus, beam B3 is passed through a quarter wave plate (QWP) with its fast axis placed at +45-degrees with respect to the y-axis. The obtained laser beam (B4) is of the form: 
\begin{equation}
\vec{E}(r,\theta)= \psi(r,\theta)\hat{e_{1}} + \exp({i\theta}) \psi(r,\theta)\hat{e_{2}}
\end{equation}
where $\hat{e_{1}}$ and $\hat{e_{2}}$ denote the right-circular and left-circular polarized basis vectors. This beam contains a lemon type C-point polarization singularity. Star type C-point polarization beam can be obtained through simple index-inversion of the lemon-type beam by passing it through a half-wave plate(HWP), shown inside dotted rectangle \cite{Pal:17}. This arm constitutes the C-point generating arm of the polarization singular generator. A beam dump (BD) is used to block the beam B2, thus the output is simply the C-point carrying beam. For generating a V-point beam, an S-waveplate (SWP) (make: Altechna) is used to convert the linear polarization into radial and azimuthal polarization. The axis of the SWP is made parallel to the y-axis of the optical setup. The laser beam (B2) is converted into either $x$-polarized or $y$-polarized beam with the help of a linear polarizer. This is now sent on the SWP. Depending on the incident beam (B4) polarization with respect to the S-waveplate's axis i.e, parallel (along $y$-axis) or perpendicular (along $x$-axis), the resulting polarization pattern is radial or azimuthal respectively. This time, a beam dump is placed in C-point generating arm.

\section{Propagation of polarization singular beam through Random phase screen}
\begin{figure}[h]
\centering
\includegraphics[width=0.8\textwidth]{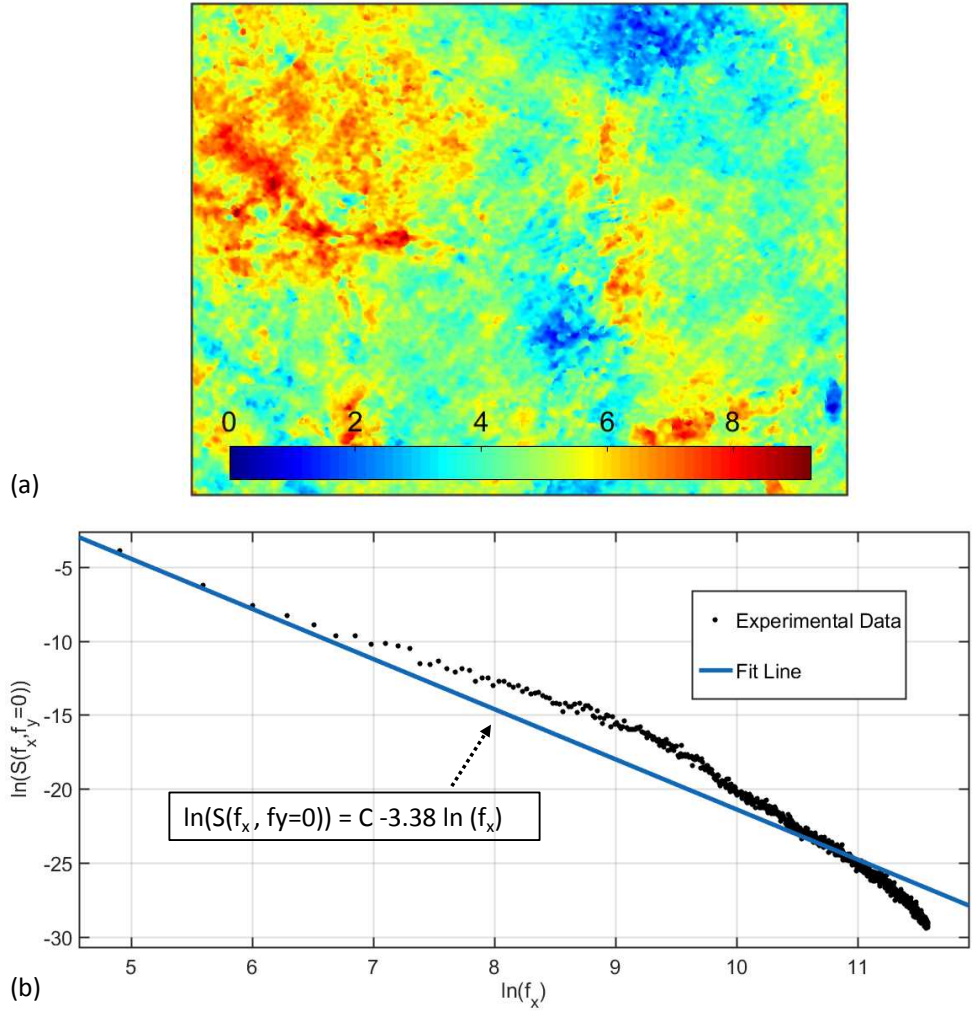}
\caption{(a) A sample random phase screen - phase recovered from DHM (b) Curve between $\ln(S(f_x,f_y=0))$ and $\ln(f_x)$}
\label{fig3}
\end{figure}
The generated polarization singular beam is passed through a random phase screen followed by a lens ($f = 20$ cm) (See Fig. \ref{fig1}(a)). The random phase screen used in the experiment was a crumpled plastic sheet. In order to see the phase structure of this sheet, its phase was measured using a Digital Holographic Microscope(DHM). Phase map of a sample random screen is shown in Fig. \ref{fig3}(a). Using $50$ such phase measurements, the phase power spectrum of the plastic sheet was obtained to be: 
\begin{equation}
S(f_x,f_y)=C(f)^{-3.38 \pm 0.05}
\end{equation} 
where $f_x$ and $f_y$ are the spatial frequencies with $f=\sqrt{{f_x}^2+{f_y}^2}$ , $k=2 \pi f$ and $C$ is a constant. The fitting was done using power law fitting tool in MATLAB. The graph of $\ln(S)$ vs $\ln(f_x)$ for the case $f_y=0$ is shown in Fig. \ref{fig3}(b). 

The Fraunhoffer diffraction pattern corresponding to the given random phase screen is observed in the back focal plane of the lens using a CCD sensor (Make: Lumnera, Model: Infinity 2-1R, pixel size $= 4.65 \mu$m)(see Fig. \ref{fig1}(a)). The lens aperture (diameter=$2.2$ mm) is chosen so that the individual speckles in the diffraction pattern are adequately sampled by CCD sensor pixels. Circular polarizers can be inserted in between to observe intensity patterns of the RCP and RCP polarization components individually.
   
\section{Results and Discussion}

\begin{figure}[h]
\includegraphics[width=0.85\textwidth]{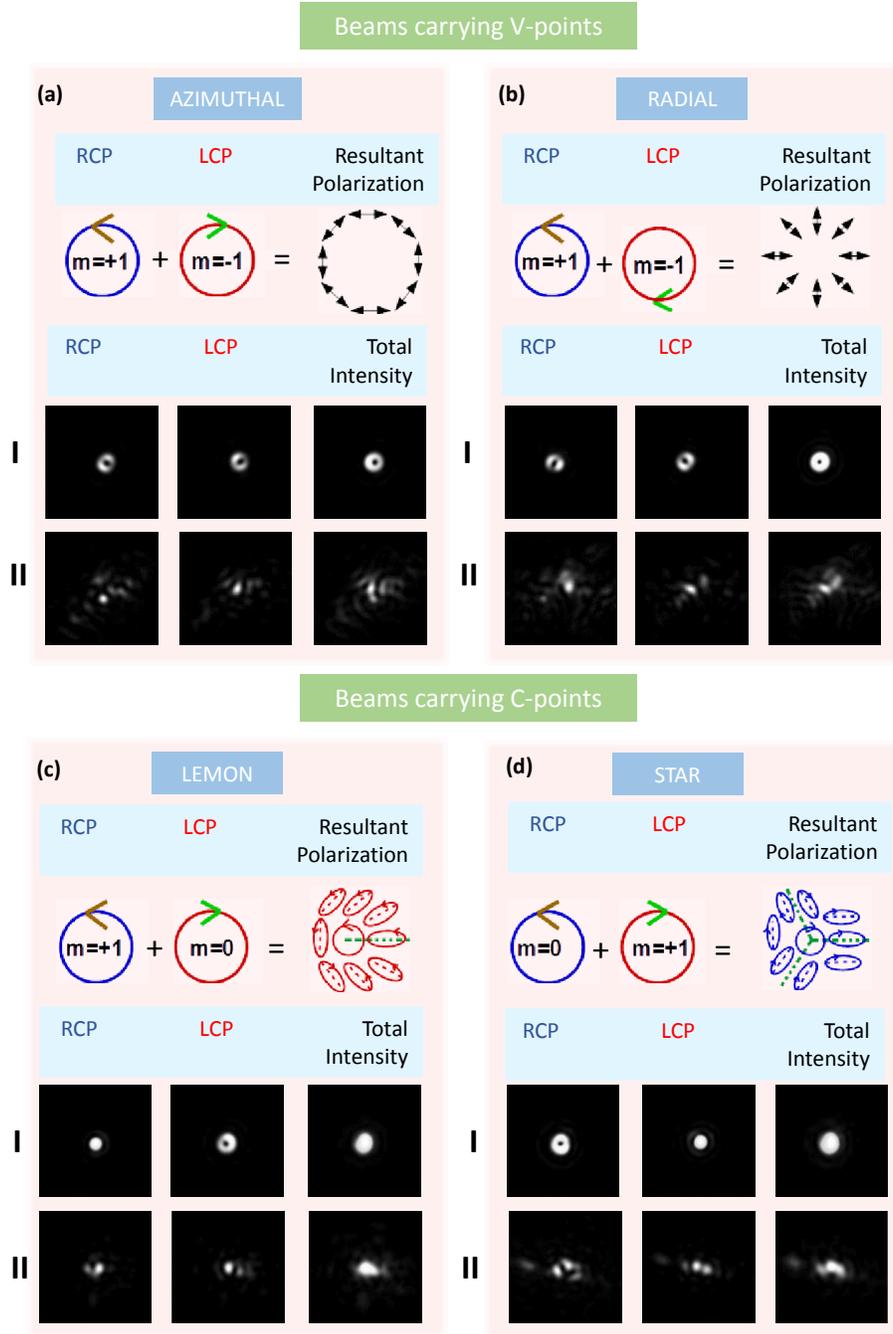}
\caption{The experimentally observed intensities for beams carrying V-point polarization singularity namely azimuthal and radial polarization and beams containing C-point polaization singularities namely lemon and star are shown in (a-b) and (c-d) respectively. The decomposition of the beams in RCP-LCP basis is also illustrated. Row I shows the Fraunhoffer diffraction intensities for no-phase screen case while row II shows the intensities when the beam has passed through a random phase screen.}
\label{fig4}
\end{figure}

Figures \ref{fig4}(a), (b) show the experimentally recorded intensities on a CCD sensor for the V-point singularity cases whereas Fig.s \ref{fig4} (c) and (d) show the experimentally recorded intensities for the C-point singularity cases. The decomposition of the V-point and C-point singularity structures in the RCP and LCP basis is illustrated diagrammatically for each of the four cases. The individual intensity patterns for the RCP and LCP states and their combined intensity profile are shown for the V-point singularities (radially and azimuthally polarized beams) and for the C-point singularities (lemon and star beams). In each case, the row {\bf I} of intensity records are for the open aperture without any random phase screen whereas the row {\bf II} of intensity records correspond to the case when a random phase screen was present in the beam path. By visual inspection we clearly observe that the intensity profile quality for the combined beam is much superior for the C-point singularity case when compared with the V-point singularity case. The correlation coefficient calculated between the experimentally recorded RCP and LCP intensities for the beams is shown in table \ref{table1}. The standard deviation in the correlation coefficient is calculated over $50$ experimental realizations of random phase screen. 
\begin{table}[h]
\centering
\caption{\bf Correlation coefficient for the LCP and RCP intensity patterns on CCD sensor for five different types of polarization singular beams.}
\begin{tabular}{ccc}
\hline
Type of Singularity \; & Singularity Structure \; & Correlation Coefficient \\
\hline
C-point & Lemon & $-0.81 \pm 0.06$ \\
C-point & Star & $-0.76\pm0.07$  \\
C-point & Lemon-star dipole & $-0.82\pm 0.06$ \\
V-point & Radial & $0.11\pm 0.41$\\
V-point & Azimuthal & $0.19 \pm 0.35$\\
\hline
\end{tabular}
\label{table1}
\end{table} 

\begin{figure}[h]
\includegraphics[width=0.85\textwidth]{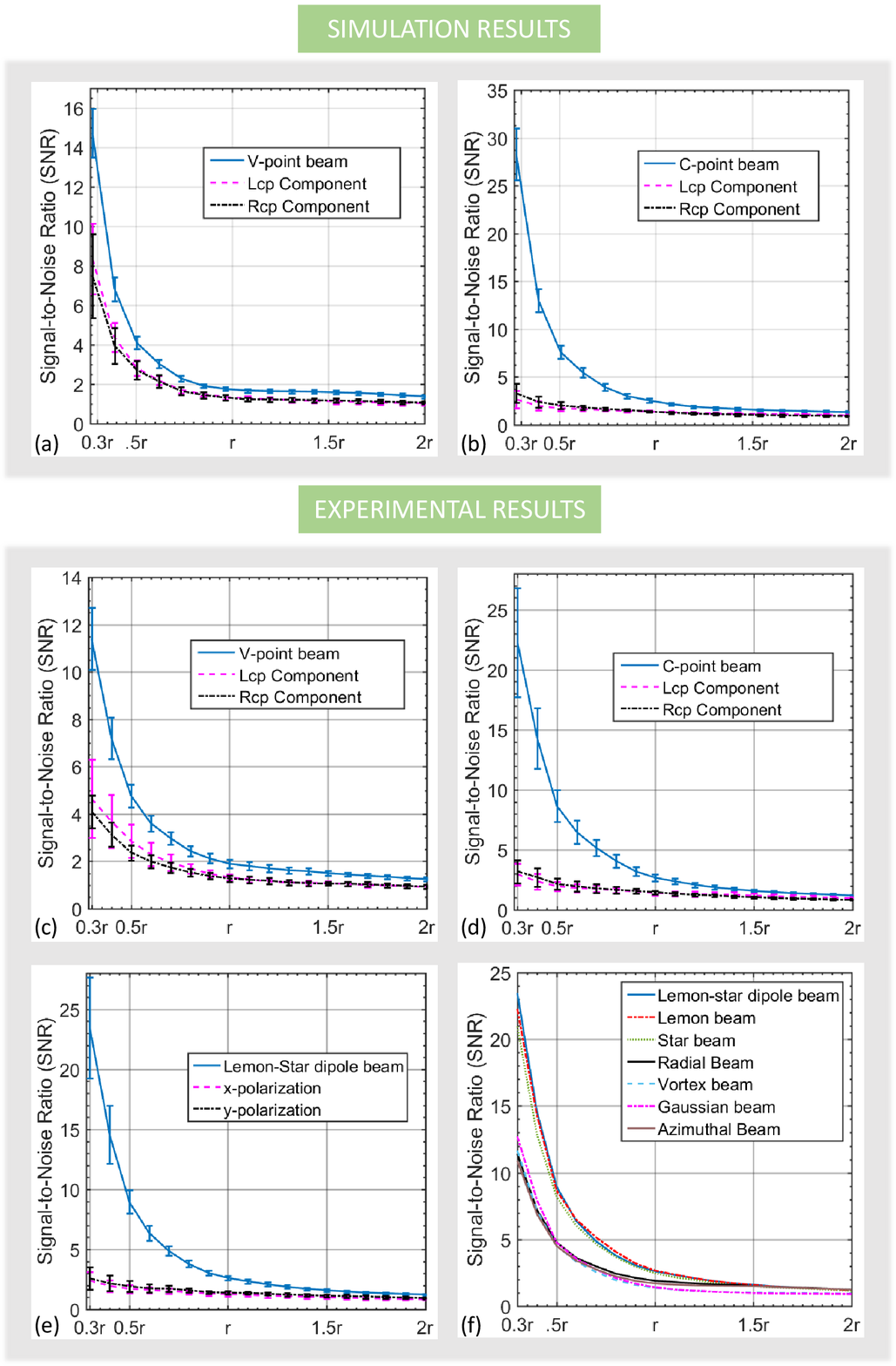}
\caption{Simulated and experimentally observed SNR curves for the total beam and its RCP-LCP components is shown for V-point and C-point beam in (a)-(c) and (b)-(d) respectively. (e) shows SNR curves for beam containing C-point dipole structure. (f) shows a summary SNR comparison for C-point carrying beams, V-point carrying beams, C-point dipole beam and homogeneously polarized Gaussian and a charge-1 Vortex beam. The parameter $r$ on the $x$-axis denotes the diffraction limited spot size.}
\label{fig5}
\end{figure}

In order to quantify the beam quality, we used the signal-to-noise ratio (SNR) of the beam intensity defined as the mean divided by the standard deviation of intensity values within a given detector area. First the peak intensity position in the detected intensity pattern for the combined beam was located and the SNR was calculated over a circle of varying radius around this peak location. The same corresponding circular regions in the observed intensity patterns were used to calculate the SNR values for the individual orthogonal polarizations. The SNR values for simulation and experimental data are plotted in Fig. \ref{fig5}(a)-(b) and Fig. \ref{fig5} (c)-(d) respectively for V-point and C-point singular beams. The plots show variation in SNR as a function of the radius of the circular region used for SNR calculation. The parameter $r$ on the $x$-axis of the plots corresponds to the radius of the diffraction limited spot. The error bars in these plots represent the standard deviation of the SNR values calculated over $50$ realizations of the random phase screen. 

In case of the V-point singularity structures, the individual polarization components (corresponding to $l = 1$ and $l = -1$ states) produce uncorrelated speckle patterns and the center of the circular region used for SNR calculations is typically near the peak of one of the patterns for $l = 1$ or $l = -1$ components. In C-point singularity case, the individual speckle patterns are negatively correlated and the center of the circle used for SNR calculation is typically not close to the peak intensities for either $l = 0$ or $l = 1$ components. The SNR values for the individual RCP and LCP components shown in Fig. \ref{fig5}(a)-(d) are therefore slightly different for the V-point and C-point cases. Figure \ref{fig5}(e) shows the experimental SNR values for a C-point singularity represented by a lemon-star dipole which is obtained by combining the $l = 0$ and $l = 1$ in $\hat{x}$ and $\hat{y}$ polarizations. The behavior of the SNR curves in Fig.s \ref{fig5}(d) and (e) is similar since they only differ in the choice of the orthogonal polarization states. The experimental SNR results for various V-point and C-point beams are summarized for convenience in Fig. \ref{fig5}(f) again to highlight that the robustness of beam to random phase fluctuation is dependent on the specific combination of OAM states rather than the orthogonal polarization basis used. From the SNR plots we observe that the SNR for the combined beam intensity profile is always more than that for individual polarization components for both V-point and C-point singularity structures. The C-point singularity beams however show approximately two-fold or more SNR gain over the V-point singularity beams when SNR is calculated over a circle of size half that of the diffraction limited spot. The complementary speckle property of the $l = 0$ and $l = 1$ states makes sure that the overall beam intensity self-corrects in presence of a fluctuating random medium such as atmospheric turbulence even when no active beam correction methods are employed. Use of beams containing C-point singularities can therefore greatly reduce the requirement of adaptive beam control which typically poses several engineering challenges in applications that need laser beam propagation through random fluctuating media. It may be worth studying combinations of higher order OAM modes in orthogonal polarization states for robustness of beam profile, however, higher order OAM beams are often unstable and split into beams with lower charge\cite{SOSKIN2001219}. The $l = 0$ and $l = 1$ is thus the most practical combination of OAM states for robust intensity profile preservation on propagation through random phase fluctuations.

\section{Conclusion}
In conclusion we have studied the effect of random phase fluctuations on the intensity profile of scalar and vector optical beams in terms of the diversity of speckle patterns produced by their individual polarization components. For the scalar beams and non-singular inhomogeneously polarized beams we observe significant positive correlation between the speckle patterns for individual polarizations. For the V-point and C-point singular beams, the speckles for individual polarization components are uncorrelated and negatively correlated respectively. The C-point singularities show optimally robust intensity profile on propagation through random phase fluctuations among all the polarization states of the beam studied here. The quality for the beams in various polarization states was quantitatively evaluated in terms of SNR of the beam intensity profile and it is observed that the C-point singularity structures show maximal SNR values suggesting their optimality for beam propagation through random phase fluctuations. Optical beams containing C-point singularity structures can be readily generated in the laboratory and their robustness as shown here has important implications in practical applications such as free-space communications and defense systems where it is necessary to deliver laser beams on a target area through turbulence. It is important to note here that the complementary speckle property associated with the orthogonal polarization components of C-point singularity structures offers an interesting passive solution to beam propagation through time-varying random media and can be considered as a general guiding principle for engineering optimally robust laser beams.

\end{document}